\documentclass[final,3p,times,twocolumn]{elsarticle}

\usepackage{amssymb}

\journal{International Workshop Belle-II and Vietnam-2017}

\begin{document}

\begin{frontmatter}


\title{High-order corrections to tauon mass in a microscopic cosmological model}


\author{Vo Van Thuan}

\address{Duy Tan University (DTU)\\
3 Quang Trung street, Hai Chau district, Danang, Vietnam\\
Vietnam Atomic Energy Institute (VINATOM)\\
59 Ly Thuong Kiet street, Hoan Kiem district, Hanoi, Vietnam
}
\ead{vvthuan@vinatom.gov.vn}
\begin{abstract}
According to our microscopic cosmological model, masses of charged leptons are induced by curvatures of hyper-spherical surfaces embedded in a 3D time-like subspace, leading to a solution of the lepton mass hierarchy problem and to a prediction of tauon mass for the first approximation. In the present study, for finest-tuning higher order approximations, there are some corrections added to higher curvatures by contributions from lower ones. In the result, the calculation in the third approximation $m_{\tau}(theor)=1776.40$ MeV fits the experimental tauon mass $m_{\tau}(exp)=1776.82(\pm.16)$ MeV  within $0.024\%$ of precision, reaching a fairly passable consistency better than $3\sigma$. On one side, it implies that for a firm consistency, a hyper-fine adjustment of calculation by some additional mechanism is needed. On the other side, our theoretical quantity demonstrates an explicit physical interpretation, which is in opposite to another theoretical calculation with a wonderful predictability by the empirical Koide formula $m_\tau(Koide)=1776.97$ MeV, i.e. within a deviation less than $1\sigma$, unfortunately, being physically unexplained. In the circumstances, a new attempt for upgrading experimental accuracy of tauon mass is desirable.
\end{abstract}


\end{frontmatter}

\section{Introduction}
Lepton mass hierarchy is a puzzle of elementary particle physics. In the standard model of particle physics (SM), masses are induced in interaction of genetic leptons  with Higgs field characterized by a global Higgs vacuum potential.  However, each kind of leptons has its own Higgs-lepton coupling constant as a free parameter of unknown origin. Generally, in the frame of SM most of efforts to solve the problem are phenomenological. In particular, Barut ~\cite{Bar1} proposed a conjecture that a magnetic self-interaction of electron induces masses of muon, tauon and heavier leptons as specific electron quantum exited states, which led roughly to their mass hierarchy, however, no heavier charged leptons were seen. Koide ~\cite{Ko1} found that the charged lepton masses are related to each other in a quite simple but unexplained empirical formula which allows to predict the mass of tauon by the experimental masses of electron and muon. Hollik and Salazar in ~\cite{Ho1} proposed for a qualitative interpretation of mass hierarchy of elementary fermions based on their mass ratios.
The situation implies that the problem of mass hierarchy is to be solved beyond SM. Following this trend, the higher dimensional general relativity would be introduced. For the modern Kaluza-Klein (K-K) models, extradimensions (EDs) are not compact, but would be extended to a macroscopic scale for inducing new physics. In the induced matter approach, Wesson and co-workers ~\cite{We1} proposed a space-time-matter theory (5D-STM) describing proper mass as a special time-like ED, which in a link with 4D space-time physics of elementary particles led to a qualitative interpretation of quantum mechanics  ~\cite{We3}. Following the induced-matter approach, our recent study ~\cite{Vo1},~\cite{Vo3} was based on the time-space symmetry with two time-like EDs which are made explicit in terms of the quantum wave function $\psi$ and the proper time variable $t_0$. There was found a duality between the quantum wave equation in 4D space-time and a relativistic geodesic description of the curved higher dimensional time-space. The problem of lepton mass hierarchy would be solved, in particular, time-like EDs for formulation of a 3D time-like configuration would be considered in a correlation with number three of charged lepton generations, in particular, tauon mass was predicted within $2.2\%$ ~\cite{Vo4} which requires a further fine-tuning to meet the experiments. In the present study, for  improving preditability of the proposed microscopic cosmological model, some corrections to the highest curvatures are added from  higher-order approximations. The article is organized as following: in Section 2, as the basis of the microscopic cosmological model, a higher dimensional bi-cylindrical general relitivity with time-space symmetry is introduced in accordance with Ref. ~\cite{Vo3}; in Section 3 the proposed model is applied to mass hierarchy of charged leptons in the first approximation (see ~\cite{Vo4}), in Sections 4 and 5 new corrections by higher-order approximations are presented.
\section{Dual solutions of gravitational equation for microscopic cosmology}
Investigation is carried out on a time-space symmetrical "lightcone" $dt_k^2=dx_l^2$ embedded in an ideal 6D flat time-space $\{t_k \mid x_l\}$, where $k,l=1 \div 3$ are summation indexes~\cite{Vo3}. Here natural units ($\hbar=c=1$) are used unless it needs an explicit quantum dimension.
For fluctuations including curved rotation and linear translation a symmetrical $\{3T,3X\}$ bi-cylindrical geometry is introduced originally. Suggesting that in both orthonormal subspaces of 3D-time and 3D-space cylindrical curvature is realized, then assuming that due to interaction of a Higgs-like potential the time-space symmetry is spontaneously broken, leading to formation of energy-momentum. In the result, the bi-cylindrical geometry is getting asymmetrical, as shown in ~\cite{Vo3}:
\begin{equation}
d\Sigma^2=(ds_0^2+ds_{ev}^2)-(d\sigma_{ev}^2+d\sigma_L^2)=dt^2-dz^2,
\label{eq5a}
\end{equation}
where: $dt^2=d\psi(t_0)^2+\psi(t_0)^2d\varphi(t_0)^2+dt_3^2$  \\
and: $dz^2=d\psi(x_n)^2+\psi(x_n)^2d\varphi(x_n)^2+dx_3^2.$
\\
The asymmetrical curved time-space is $\{3T,3X\} \equiv \{\psi(t_0),\varphi(t_0),t_3 \mid \psi(x_n),\varphi(x_n),x_3\}$. The time-like and space-like intervals in Geometry $(\ref{eq5a})$ separate into even ($ds_{ev}$ and $d\sigma_{ev}$) and odd ($ds_0$ and $d\sigma_L$) constituents, which means the corresponding cylindrical accelerations can flip for and back  (as an even-term) or can not flip (as an odd-term)  in relation to the cylindrical axis.
In observing an individual fermion elementary particle, e.g. a free lepton with (pseudo-)spins $\vec {\tau}$ and $\vec s$, its projections $\tau_k$ or $s_l = \pm 1/2$ can be fixed on the longitudinal axes of $\{t_k\}$ and $\{x_l\}$, respectively, leading to a cylindrical dynamical model. Being embedded in 6D-lightcone, cylindrical variables $\{\psi,\varphi \}$ are getting functions of linear coordinates $\{t_k,x_l\}$ and two 3D-local affine parameters $t_0$ and $x_n$ which are introduced in according to projection of (pseudo-)spins $\vec {\tau}$ and $\vec s$, respectively.
\\
Applying the bi-cylindrical geometry to orthonormal subspaces 3D-time and 3D-space, the corresponding gravitational equation in an absolute  $\{3T,3X\}$-vacuum reads:
\begin{equation}
R^m_i-\frac{1}{2}\delta^m_i R=0,
\label{eq7}
\end{equation}
which leads to an equivalent $\{3T,3X\}$- Ricci vacuum equation $R^m_i=0$. As $\psi=\psi(y)$ and  $\varphi=\varphi(y)$, it is assumed that the Hubble law of the cosmological expansion is applied for the bi-cylindrical model of microscopic space-time:
$\frac{\partial \psi}{\partial y} =v_y=H_y\psi$. Therefore: $\left[\frac{\partial y}{\partial \psi}\right]=\frac{1}{H_y \psi}$, where $v_y$ is expansion rate proportional to  the "microscopic scale factor" $\psi$ and $H_y$  is a "microscopic Hubble constant"; $y\equiv\{t,z\}\equiv\{t_0,t_3,x_n,x_3\}\in \{t_i,x_j\}$. As $\{i,j\}$ are summation indexes of curved coordinates then $\{t_i\}$ and $\{x_j\}$ are explicitly embedded in 3D-time or in 3D-space, correspondingly.
Equation $R^\psi_\psi=0$ is a solution of $(\ref{eq7})$, in which the acceleration term in 3D-time  is enhanced strongly due to interaction with a Higgs-like potential $V_T$ of a time-like "cosmological constant" $\Lambda_T$, leading to an asymmetrical bi-geodesic equation:
\begin{equation}
\frac{\partial^2 \psi}{\partial t^2}-\frac{\partial^2 \psi}{\partial {x_j}^2}=\left [\Lambda_T -\left(\frac{\partial \varphi}{\partial x_n}\right)_{even}^2 -\Lambda_L \right ]\psi,
\label{eq21}
\end{equation}
where $\Lambda_L\equiv \left(\frac{\partial \varphi}{\partial x_n^L}\right)^2$ is a small space-like P-odd "cosmological constant" caused by the global weak interaction leading to the left-handed space. Being originated from the higher dimensional gravitational equation $(\ref{eq7})$, Equation $(\ref{eq21})$ describes the microscopic cosmological geodesic evolution of time-space curvatures by its monotone exponential solution $\psi=\psi_0 e^{\pm \varphi}=\psi_0 e^{\pm (\Omega t+k_j x_j)}$.
\\
On the other side, as variables  $\{y\}$ can reform as well as $\{y\}\leftrightarrow\{iy\}$ then Equation $(\ref{eq21})$ leads to a wave-like  representation with $\psi_w\equiv\psi(y\rightarrow iy)\sim e^{i\varphi}=e^{i(\Omega t-k_j x_j)}$, namely:
\begin{eqnarray}
&& -\frac{\partial^2 \psi}{\partial t^2}+\frac{\partial^2 \psi}{\partial {x_j}^2}= \nonumber \\
&& =\left [\left(\frac{\partial \varphi}{\partial t_0^+}\right)^2-B_e ( k_n.\mu_e)_{even}^2 -\left(\frac{\partial \varphi}{\partial x_n^L}\right)^2 \right ]\psi,
\label{eq22a}
\end{eqnarray}
where $B_e$ is a calibration factor  and $\mu_e$ is magnetic dipole moment of charged lepton, being P-even value as its orientation correlates with spin vector $\vec s$. The transformation from the exponential solution to the wave-like one is realized by replacing variables: $t\rightarrow -it$ and $x_j \rightarrow ix_j$, as well as of their corresponding covariant derivatives. This procedure is equivalent to transformation from an external observation to an internal investigation in the phase frame ~\cite{Vo3}.  Rescaling with Planck constant by quantum dynamical operators $\frac{\partial }{\partial t}\rightarrow \hat{E}=i\hbar\frac{\partial}{\partial t}$; $\frac{\partial }{\partial x_j} \rightarrow \hat{p_j}=-i\hbar\frac{\partial }{\partial x_j}$ and making the functional parameter $\psi$ of a scale of Compton length, a generalized Klein-Gordon-Fock equation is formulated from Representation $(\ref{eq22a})$ as:
\begin{equation}
-\hbar^2\frac{\partial^2 \psi}{\partial t^2}+\hbar^2\frac{\partial^2 \psi}{\partial x_j^2}-m^2\psi=0,
\label{eq23a}
\end{equation}
where the square mass term $m$ consists of the following components:   $m^2=m_0^2-m_s^2-m_L^2$. Except the rest mass $m_0$, the P-even contribution $m_s$ is linked with an external curvature of spinning in 3D-space  and a small non-zero mass factor $m_L\ll m_s$ which proves a tiny internal curvature of our realistic 3D-space. In general, Equation $(\ref{eq23a})$ is reminiscent of the squared Dirac equation of a free charged lepton.
In duality to the wave-like equation $(\ref{eq23a})$, the geodesic equation $(\ref{eq21})$ in a homogeneity condition (i.e. without the translational terms) being equivalent to de Sitter-like solutions can serve for modeling Hubble expansion in the microscopic time-space. In analogue to the standard model of macroscopic cosmology, a microscopic cosmological model is proposed to solve the mass hierarchy problem of leptons ~\cite{Vo4}.
\section{Charged lepton mass hierarchy in the first order approximation}
In 4D space-time we assume that all charged leptons are to involve in the same basic time-like cylindrical geodesic evolution with an internal curvature of the time-like circle $S_1(\varphi^+)$, where $\varphi^+=\varphi({t_0}^+)$ is azimuth rotation in the plane $\{t_1,t_2\}$ about $t_3$ and the sign $+$ means an evolution toward the future (see ~\cite {Vo4}). This universal feature determines the common properties of all charged lepton generations, except their mass hierarchy. Developing higher orders of curvature, we consider a generalized 3D time-like spherical system, described by nautical angles $\{\varphi^+,\theta_T,\gamma_T\}$, where $\theta_T$ is a zenith in the plane $\{t_1,t_3\}$ and $\gamma_T$ is another zenith in the orthogonal plane $\{t_2,t_3\}$. Coexisting in the same time-like cylindrical evolution $\varphi^+$, 4D observers see electron oscillating along a line-segment of the time-like amplitude $\Phi$, formulating one-dimensional comoving "volume": $V_1(\varphi^+)=\Phi=\psi.T$; where $T$ is the time-like Lagrange radius.
In similar to the standard cosmological model, introducing a so-called microscopic cosmological model to the 3D time-like sphere, based on Equation $(\ref{eq21})$ in a homogeneity condition, we consider $\Phi$ as the time-like microscopic Hubble radius and the functional parameter $\psi$ as the time-like scale factor.
The highest order curvatures $C_n$ of $n-$hyper spherical surfaces are inversely proportional to $n-$ power of the time-like scale factor as $C_n\sim \psi^{-n}$. In particular, the energy density of electron correlates with its internal curvature as: $\rho_1=\epsilon_0/\psi$; where $\epsilon_0$ is assumed as a universal lepton energy factor. The mass of electron is determined as:
\begin{equation}
m_1=\rho_1V_1=\rho_1\Phi=\epsilon_0.T=\epsilon_0 W_1.
\label{eq22}
\end{equation}
The value $W_1=T$ is the time-like Lagrange "volume" of electron. For muon and tauon except the common time-like cylindrical curved evolution $\varphi^+$, our 4D-observers can see some additional extradimensional curvatures come from simplest configurations of hyper-spherical surfaces $S_1(\theta_T)$ and $S_1(\gamma_T)$ or $S_2(\theta_T,\gamma_T)$. Those additional curvatures are external to 4D-observers as they are not involved in, then they can recognize the corresponding hyper-surfaces as the foot-prints being projected on the basic cylindrical evolutional axis $dt$. For this reason, the additional higher dimensional curvatures are to be seen with a fixed maximal amplitude $\Phi$ and the corresponding time-like "comoving volumes" $V_n(\Phi)$ are calculated as follows:
\begin{equation}
 V_n(\Phi)=\int_0^\Phi S_{n-1}(v)dv=\Phi.S_{n-1}(\Phi)=V_1S_{n-1}.
\label{eq23}
\end{equation}
The energy distribution of n-hyper spherical configuration relates to electron density as: $\rho_n=\rho_1/\psi^{n-1}$. Therefore, the mass corresponding to $n-$dimensional configuration reads:
\begin{eqnarray}
m_n=\rho_n.V_n(\Phi)&&=(\rho_1/\psi^{n-1})V_1S_{n-1}= \nonumber \\
&&=W_1\rho_{n-1}S_{n-1}.
\label{eq24}
\end{eqnarray}
For homogeneity condition of motion equation of particle at rest, the simplest additional $S_1$ configuration is: $[S_1(\theta_T)+S_1(\gamma_T)]$, then the lepton mass of 2D time-like curved particle is:
\begin{equation}
m_2=W_1 \rho_1[S_1(\theta_T)+S_1(\gamma_T)]=\epsilon_04\pi.T^2=\epsilon_0 W_2.
\label{eq25}
\end{equation}
For the simplest additional $S_2(\theta_T,\gamma_T)$ configuration the lepton mass of 3D time-like curved particle is:
\begin{equation}
m_3=W_1 \rho_2 S_2(\theta_T,\gamma_T)=\epsilon_04\pi.T^3=\epsilon_0 W_3.
\label{eq26}
\end{equation}
There in $(\ref{eq25})$ or $(\ref{eq26})$ $W_n$ is dimensionless Lagrange volume. The equations of lepton mass are obtained here in the first approximation only, because the time-like curvatures of hyper-spherical surfaces would contain more precise terms. There are in $(\ref{eq22})$, $(\ref{eq25})$ and $(\ref{eq26})$ two free parameters: the lepton energy factor $\epsilon_0$ and the time-like Lagrange radius $T$ which would be determined, in principle, by experimental masses of two from three charged leptons and to use for predicting the mass of the third lepton. In particular, according to $(\ref{eq22})$ and $(\ref{eq25})$  using the experimental masses of electron and muon for calibration, we found the lepton energy factor $\epsilon_0=31.05602942$ keV and $T=16.45409724$. Now Equation $(\ref{eq26})$ allows to predict the absolute mass of tauon (in MeV) in a mass hierarchy, as follows:
\begin{eqnarray}
&&m_1:m_2:m_3=\nonumber \\ 
&&=0.510998928:105.6583715:1738.51.
\label{eq27}
\end{eqnarray}
It is to compare with the experimental data of charged lepton masses from ~\cite{Be1}:
\begin{eqnarray}
&&m_e:m_\mu:m_\tau= 0.510998928(11): \nonumber \\
&&:105.6583715(35):1776.82 (16).
\label{eq28}
\end{eqnarray}
There the predicted tauon mass is in a good consistency with experimental data, within $2.2\%$ of relative deviation, even for the first order of approximation. 
\section{The second approximation of tauon mass by minor curvatures}
Let us search for fine-tuning tau mass by contribution from minor curvatures $C_k$ to the major curvature $C_n$ producing lepton mass $m_n$. This implies that $S_1$ minor curvature is to be added to $S_2$ major curvature, while $S_1$ and $S_2$ minor curvatures should be added to the $S_3$ major curvature. Formula of electron mass is now rewritten in the second order approximation as:
\begin{equation}
m_1(2)=m_1(T_2)=\epsilon_2.T_2.
\label{eq22A2}
\end{equation}
Formula of muon mass is upgraded as:
\begin{equation}
m_2(2)=m_2(T_2)\left[1+\delta \left(\frac{C_1}{C_2}\right)\right],
\label{eq25A1}
\end{equation}
where $m_2(T_2)=\epsilon_2 4\pi T_2^2$; $\delta\left(\frac{a}{b}\right)$ implies a quantity of the order of ratio $a/b$. As the curvatures $C_1$ and $C_2$ are of different dimensions, they are normalized by corresponding dimensionless Lagrange volumes as follows:
\begin{eqnarray}
&&m_2(2)=m_2(T_2)\left[1+\frac{W_1}{W_2}\right]= \nonumber \\
&&m_2(T_2)\left[1+\frac{m_1(T_2)}{m_2(T_2)}\right]=m_2(T_2)+m_1(T_2).
\label{eq25A2}
\end{eqnarray}
In similar, Formula of tauon mass will have corrections upto the order of minor curvature $C_2$ as:
\begin{equation}
m_3(2)=m_3(T_2)\left[1+\delta \left(\frac{C_1}{C_3}\right)+\delta \left(\frac{C_2}{C_3}\right)\right],
\label{eq26A1}
\end{equation}
which leads to: 
\begin{eqnarray}
m_3(2)&&=m_3(T_2)\left[1+\frac{m_1(T_2)}{m_3(T_2)}+\frac{m_2(T_2)}{2m_3(T_2)}\right]= \nonumber \\
&&=m_3(T_2)+m_1(T_2)+\frac{1}{2} m_2(T_2),
\label{eq26A2}
\end{eqnarray}
where $m_3(T_2)=\epsilon_2 4 \pi T_2^3$; the factor of $1/2$ for $m_2$ implies that the principal muon mass $(\ref{eq25})$ consists of double curvature of $S_2$. The different factors of $C_2$ contribution for muon and tauon mean that in Equation  $(\ref{eq25A1})$ $C_2$ refers to muon mass, while in Equation $(\ref{eq26A1})$ $C_2$ relates to a correction to tauon mass, taking a single $C_2$ only. In the result, both corrected configurations of muon and tauon contain equally a structural term $m_1(T_2)$ to meet the requirement that they are involved in the same basic time-like cylindrical geodesic evolution like electron.\\
In Equations  $(\ref{eq22A2})$ and  $(\ref{eq25A2})$ two new free parameters $T_2$ and $\epsilon_2$ may be determined based on experimental electron and muon masses as follows: 
\begin{equation}
T_2=\frac{1}{4 \pi}(R_{21}-1)=16.37451977.
\label{eqT2}
\end{equation}
and $\epsilon_2=31.20695661$ (keV); where $R_{21}$ is the mass ratio of muon to electron. Now Equation $(\ref{eq26A2})$ for calculation of tauon mass in the second approximation leads to: $m_3(2)=1774.82$ (MeV). The uncertainty of this theoretical prediction is ignorable, as it depends only on experimental errors of electron and muon masses, being far less than one of tauon.  The calculation in the second order approximation deviates from the experimental tauon mass by $0.11\%$, which is by 20 times better than the prediction by $(\ref{eq26})$ in the first approximation.

\section{The third approximation by finest tuning minor curvatures}
Let add the next corrections to minor curvatures $C_k$ upto n-order. Formula of electron mass is modified as:
\begin{equation}
m_1(n)=m_1(T_n)=\epsilon_n.T_n.
\label{eq22A3}
\end{equation}
Formula of muon mass is upgraded as:
\begin{eqnarray}
m_2(n)&&=m_2(T_n)\left[1+\sum_{k=1}^n \delta \left(\frac{C_1}{C_2}\right)^k\right]= \nonumber \\
&&=m_2(T_n)\sum_{k=0}^n \delta \left(\frac{C_1}{C_2}\right)^k.
\label{eq25A31}
\end{eqnarray}
After normalization it leads to:
\begin{eqnarray}
m_2(n)&&=m_2(T_n) \sum_{k=0}^n \left[\frac{m_1(T_n)}{m_2(T_n)}\right]^k= \nonumber \\
&&=m_2(T_n)+m_1(T_n)\sum_{k=0}^n \left[\frac{m_1(T_n)}{m_2(T_n)} \right]^k.
\label{eq25A32}
\end{eqnarray}
In similar, Formula of tauon mass will have corrections upto the order of minor curvatures $C_2$ as:
\begin{eqnarray}
&&m_3(n)=m_3(T_n)+m_1(T_n) \sum_{p=0}^n \delta \left(\frac{C_1}{C_2}\right)^p. \nonumber \\
&&.\sum_{k=0}^n \delta \left(\frac{C_1}{C_3}\right)^k+\frac{1}{2} m_2(T_n)\sum_{k=0}^n \delta \left(\frac{C_2}{C_3}\right)^k,
\label{eq26A31}
\end{eqnarray}
which leads to: 
\begin{eqnarray}
&&m_3(n)=m_3(T_n)+m_1(T_n)\sum_{p=0}^n \left[\frac{m_1(T_n)}{m_2(T_n)}\right]^p. \nonumber \\
&&\sum_{k=0}^n \left[\frac{m_1(T_n)}{m_3(T_n)}\right]^k+\frac{1}{2}m_2(T_n) \sum_{k=0}^n \left[\frac{m_2(T_n)}{2m_3(T_n)}\right]^k.
\label{eq26A32}
\end{eqnarray}
When $n\rightarrow \infty$ the summations converge as:
\begin{equation}
\sum_{k=0}^{\infty} \frac{1}{\rho_{ij}^k}=\frac{\rho_{ij}}{\rho_{ij}-1},
\label{eq42ij}
\end{equation} 
where for $i>j$: $\rho_{ij}=\frac{m_i(T_{\infty})}{m_j(T_{\infty})}>1$. In the result, the corresponding masses for $n\rightarrow \infty$ converge to finite quantities. For electron mass it is modified as:
\begin{equation}
m_1(\infty)=m_1(T_{\infty})=\epsilon_{\infty}.T_{\infty}.
\label{eq22A3f}
\end{equation}
Formula of muon mass is upgraded as:
\begin{equation}
m_2(\infty)=m_2(T_{\infty})+m_1(T_{\infty}) \frac{\rho_{21}}{\rho_{21}-1},
\label{eq25A3f}
\end{equation}
where $m_2(T_{\infty})=\epsilon_{n} 4 \pi T_{n}^2=\epsilon_{\infty} 4 \pi T_{\infty}^2$. 
In similar, Formula of tauon mass will have corrections upto the order of minor curvatures $C_2$ as:
\begin{eqnarray}
m_3(\infty)&=&m_3(T_{\infty})+m_1(T_{\infty})\frac{\rho_{21}}{\rho_{21}-1}\frac{\rho_{31}}{\rho_{31}-1}+ \nonumber \\
&& +\frac{1}{2}m_2(T_{\infty})\frac{2\rho_{32}}{2\rho_{32}-1}.
\label{eq26A3f}
\end{eqnarray}
where $m_3(T_{\infty})=\epsilon_n 4 \pi T_n^3=\epsilon_{\infty} 4 \pi T_{\infty}^3$. \\
In Equations  $(\ref{eq22A3f})$ and  $(\ref{eq25A3f})$ two new free parameters $T_{\infty}$ and $\epsilon_{\infty}$ may be determined based on experimental electron and muon masses as follows: 
\begin{equation}
T_{\infty}=\frac{1}{4 \pi} \rho_{21}=16.37413114.
\label{eqTf}
\end{equation}
and $\epsilon_{\infty}=31.20769729$ (keV).
Now Equation $(\ref{eq26A3f})$  leads to tauon mass in the third approximation as: $m_3(\infty)=1776.40$ (MeV). This theoretical prediction, as $m_3(2)$ in Equation $(\ref{eq26A2})$ for the previous approximation, has ignorable uncertainty due to high precision of the experimental electron and muon masses. The calculation in the third order approximation is by more than 90 times better than the prediction by Equation $(\ref{eq26})$ in the first approximation. 
\section{Discussion and Conclusions}
Prediction of tauon mass by Koide formula based on electron and muon masses, leads to the quantity $m_{\tau}(Koide)=1776.97$ MeV being in a wonderful agreement with experimental tauon mass by a deviation less than $1\sigma$. A geometrical interpretation of Koide formula was assumed in ~\cite{Koc1} where mass correlations are expressed through Descartes-like circles or with their corresponding squared curvatures. However, no more physics could be developed after all. In opposite, our microscopic cosmological model with the above fine-tuning calculation of tauon mass leads to another prediction with a fairly passable consistency within $3\sigma$, but this latest calculation demonstrates an explicit physical interpretation, which may serve a solution to the long-standing problem of charged lepton mass hierarchy by implementing a hierarchy of higher-order curvatures. Naturally, the time-space symmetry leads to accepting a 3D-time concept, the dimension of which correlates strictly with the number of lepton generations. As our corrected calculation $m_\tau(\infty)=1776.40$ MeV still deviates from the experimental tauon mass by $2.65\sigma$, it needs further searching for any other mechanism for hyper-fine correction of the theory. Despite this, in the circumstances, when those two independent theoretical predictions are approaching to the experimental tauon mass from opposite sides, new attempts for upgrading the experimental accuracy of tauon mass by 2-3 times better would be desirable.
\section*{Acknowledgment}
The author thanks deeply N.B. Nguyen (Thang Long University) for technical assistance.





\end{document}